\documentclass[aps,nanotechnology,twocolumn,groupedaddress]{revtex4}
\usepackage[dvips]{graphicx}
\usepackage{amssymb,amsmath}
\usepackage{dcolumn}
\usepackage{bm}
\usepackage[usenames]{color}
\usepackage{subfigure}

\begin{document}

\title{ Nano-Hall sensors with granular Co-C}
\author{Mihai Gabureac}
 \email[Electronic mail: ]{gabureac@gmail.com}
 \author{Laurent Bernau}
 \author{Ivo Utke}
\affiliation{EMPA, Laboratory for Mechanics of Materials and Nanostructures,
Feuerwerkerstrasse 39, 3602 Thun, Switzerland}
 \homepage{http://www.empa.ch/abt128}
\author{Giovanni Boero}
\affiliation{Swiss Federal Institute of Technology at Lausanne (EPFL), CH-1015 Lausanne, Switzerland}

\date{\today}

\begin{abstract}
We analyzed the performance of Hall sensors with different Co-C ratios, deposited directly in nano-structured form, using Co$_2$(CO)$_8$ gas molecules, by focused electron or ion beam induced deposition. Due to the enhanced inter-grain scattering in these granular wires, the Extraordinary Hall Effect can be increased by two orders of magnitude with respect to pure Co, up to a current sensitivity of 1 $\Omega/T$. We show that the best magnetic field resolution at room temperature is obtained for Co ratios between 60\% and 70\% and is better than  1 $\mu$T/Hz$^{1/2}$. For an active area of the sensor of 200 $\times$ 200 nm$^2$, the room temperature magnetic flux resolution is $\phi_{min} = 2\times10^{-5}\phi_0$, in the thermal noise frequency range, i.e. above 100 kHz.
\end{abstract}


\maketitle

\section{Introduction}

Nano-sized magnetic sensors are needed for the detection of spatially inhomogeneous magnetic fields produced by dipolar sources such as the magnetic beads used in medicine and biology \cite{Sandhu} or the magnetic grains used in magnetic recording \cite{Freitas-MRsensors}, as well as in other applications where a high spatial resolution is essential \cite{Dolabdjian,Novoselov,Matveev-nanoHall}.

Local vapor ``deposition induced by focused electron or ion beams''  (FEBID or FIBID) is a well established  technique \cite{Ivo-ReviewFebid} that allows the deposition of Hall sensors directly in nano-structured form without need for resist lithography and lift-off processes. Gas molecules are delivered inside the microscope chamber to the substrate surface where they reversibly adsorb. Their local dissociation by the focused electron or ion beam will result in a nonvolatile deposit while volatile reaction products are pumped away. The non-thermal dissociation of Co$_2$(CO)$_8$ by electrons or ions results in a co-deposition of carbon. The deposited material is granular consisting of Co-nanocrystals embedded in an amorphous carbonaceous matrix: Co$_x$(CO)$_{1-x}$. Low Co ratios $x$ can reduce the electrical transport \cite{Lau}, while high purity deposits have shown properties close to pure Co \cite{deTeresa-highpurity}.

Previously, sub-micron sized Co-FEBID Hall sensors have been shown to have an enhanced Hall sensitivity $S_I = 1 \Omega/T$ \cite{Giovanni-FebidHall} with respect to similarly sized pure Co sensors grown by e-beam evaporation in UHV, making them a good candidate for detecting low magnetic fields. In this article we investigate the characteristics of Co-C FEBID and FIBID Hall sensors with variable Co ratios ($0.5<x<0.8$) and sizes in the 100 nm range.

In a ferromagnetic material, the Hall resistivity is given by the sum of the ordinary Hall effect (OHE), generated by the Lorentz force acting on the charge carriers, and the extraordinary Hall effect (EHE), proportional to the spontaneous magnetization \cite{Berger,Fert}:
\begin{eqnarray}
\rho_\mathrm{H} = \rho_\mathrm{OH} + \rho_\mathrm{EHE} = \mu_0 (R_\mathrm{0} H + R_\mathrm{S} M_z)
\label{eq1}
\end{eqnarray}
where H is the applied magnetic field, perpendicular to the Hall cross, $M_z$ is the spontaneous magnetization along the field direction z, $R_0$ and $R_S$ are the ordinary and spontaneous Hall constants, and $\mu_0$ is the free space permeability.

Because the OHE is inversely proportional to the charge density $n$ ($R_0 = (n e)^{-1}$), in a metal, the OHE is negligible compared to the EHE, and the working field range is limited by the saturation field. The Hall signal is then providing a local measurement of the magnetization $M_z(H)$ 
in the nanostructured cross. The EHE resistivity $\rho_{EHE}$ is linked to the longitudinal resistivity $\rho$ due to the skew and side-jump scattering mechanisms: $\rho_\mathrm{EHE} = \alpha \rho + \beta \rho^2$. Thus, the EHE depends on the spontaneous magnetization and can be enhanced by surface scattering, if the film thickness becomes smaller than the mean free path.

TEM characterization of Co-FEBID tips \cite{Ivo-TEM, Ivo-Conanocrystals} showed 
that the deposited material is granular.
This nano-composite structure is similar to that of co-sputtered heterogenous granular magnetic films embedded in an insulating matrix. According to the domain wall theory (DW), below 10 nm (the exchange length in Co), the nanoparticles are single domains \cite{Park} forming films which are superparamagnetic at room temperature \cite{Denardin}.

The transport in this granular material is influenced by the intergrain scattering with the resistivity depending not only on the nanoparticles  
distribution but also on their sizes (radius $r$), because of the surface scattering being proportional to the interfacial area per unit volume: $\rho \sim 1/r$. This also leads to an increase in the EHE \cite{Brouers}, known as the giant Hall effect (GHE) \cite{Socolovsky} when the metal concentration is approaching the percolation threshold $x_C$\cite{PAKHOMOV}.


\section{Experimental}

 \begin{figure}
        \mbox{\includegraphics[width=3.2in]{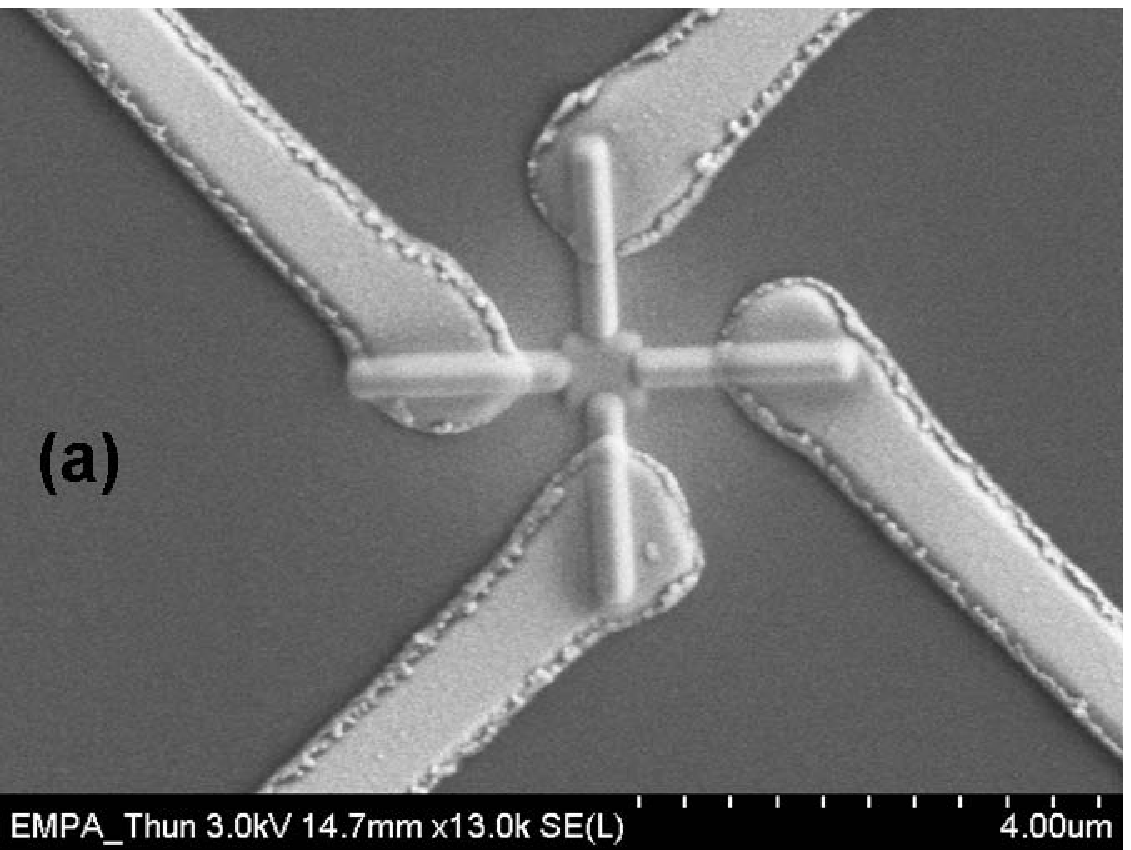}}
        \mbox{\includegraphics[width=3.2in]{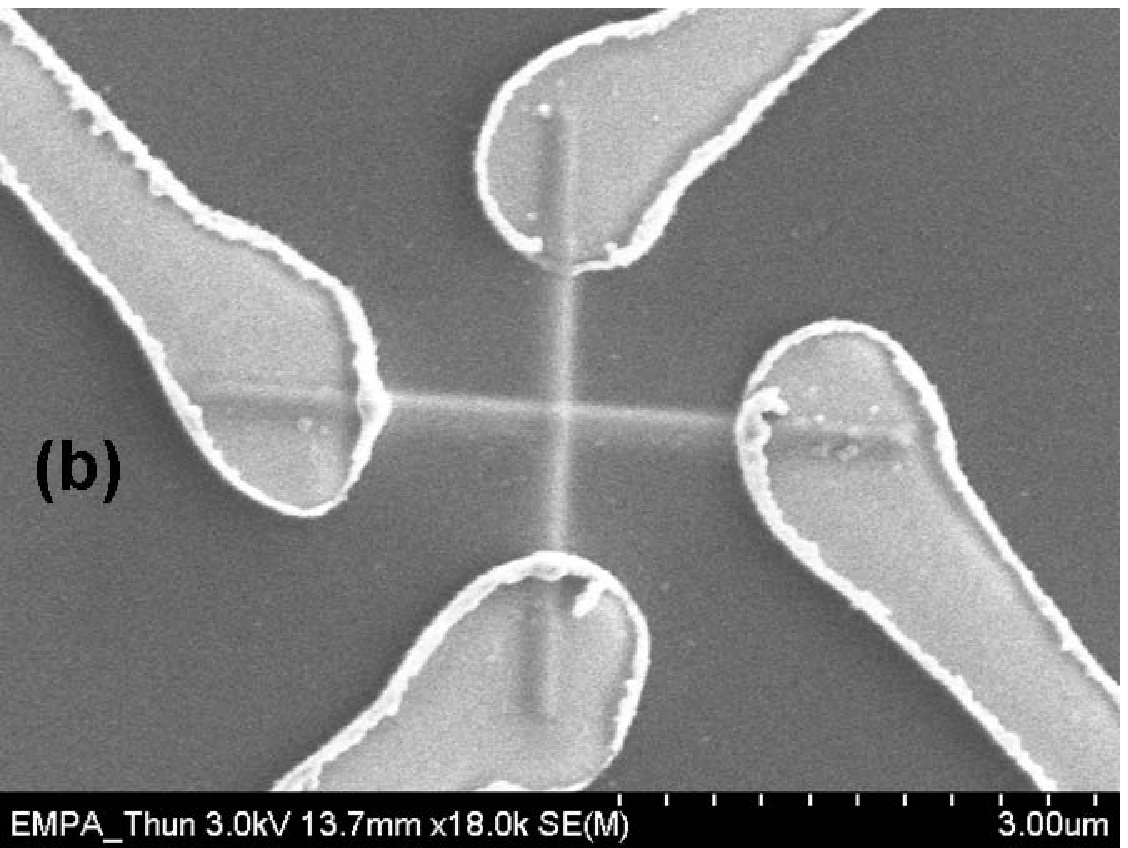}}
  \caption{Hall sensors with different designs fabricated by FEBID with CO$_2$(CO)$_8$: (a) thin active area, (b) uniform thickness}
  \label{Fig1}
  \end{figure}

We have used FEBID and FIBID to fabricate Hall sensors with thicknesses ($t$) between a few tens up to a few hundreds nanometers and widths ($w$) between 200 and 500 nm. The Co-C sensors were deposited on top of  pre-defined Cr-Au electrodes,  obtained by standard UV and e-beam lithography, with thicknesses around 100 nm and a gap ($l$) between electrodes below $2 \ \mathrm{\mu m}$ on Si wafers with an oxide layer of 200 nm. For sensors with thicknesses smaller than the Cr/Au electrode, in order to avoid step coverage problems, we have defined Hall pads with a thickness higher than the central active area contributing to the EHE effect, as can be seen in Fig \ref{Fig1}a, while for thicker sensors we have used a uniform deposition, as can be seen in Fig \ref{Fig1}b. The composition of the FEBID nanostructure can be tuned by varying 
the precursor flow, beam current, deposition pressure, dose per pixel (dwell time) and refreshment rate. The typical gas flux for the Co$_2$(CO)$_8$ precursor (argon stabilized, from VWR International, Dietikon, Switzerland) delivered to the substrate was $5.7 \times 10^2$ monolayers/s. More details are given in  \cite{Laurent}. Ex situ EDX analysis of the Co-C ratio in these deposits showed a large composition window, which was also confirmed by the resistance measurements.

\section{Hall characterization}
 \begin{figure}
 \includegraphics[width=3.8in]{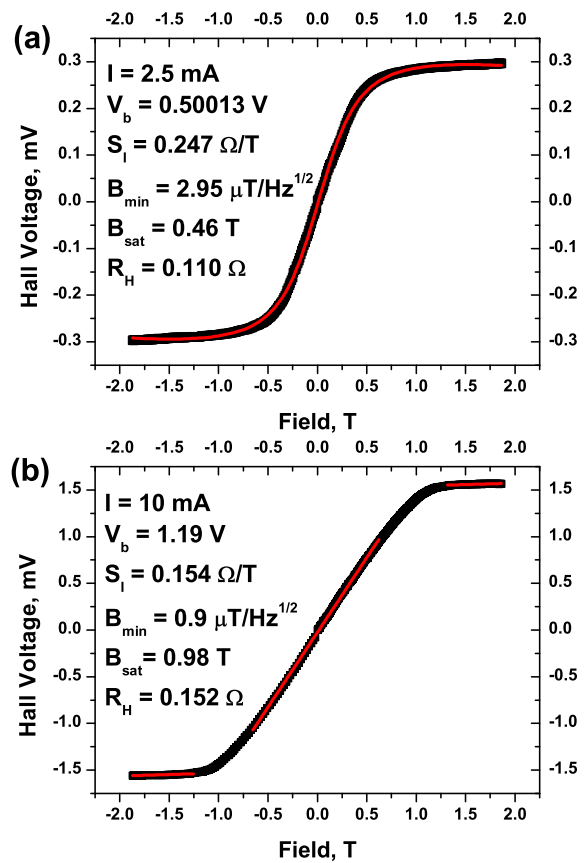}%
 \caption{ 
 Hall voltage as a function of the applied magnetic field for: (a) typical FEBID sensor, fitted with the Langevin Equation (\ref{Langevin}), (b) typical FIBID sensor, with linear fits at saturation (OH) and about zero field (EHE) as in Equation (\ref{eq1})\label{Fig2}}.
 \end{figure}
 By choosing  different sets of parameters for the FEBID/FIBID, it is possible to adjust the nanoparticles sizes and their relative distance, which determine their magnetic characteristics (saturation field and sensitivity- see Fig \ref{Fig3},\ref{Fig4}). In order to differentiate between deposits with similar Co atomic ratios (Fig \ref{Fig3}a), EDX analysis is not sufficient and we have used the Hall measurements as a local characterization method for monitoring the magnetization reversal in the active area of the Co-C sensor.

 \subsection{Langevin superparamagnetism}

 The typical magnetic field dependence of the Hall Voltage $V_\mathrm{H}$ is plotted in Fig \ref{Fig2}. The Hall voltage was measured in a homogeneous magnetic field perpendicular to the sensor, at room temperature. If interactions between particles are negligible, one can use the Langevin theory for paramagnetism which describes the variation of the spontaneous magnetization with the applied field $H$:  $ M_z = M_S L(\mu H/ k_B T)$ where $L(x)=\coth x -1/x$, with $x= \mu H/ k_B T = a B$, by replacing $M_z$ in Eq. \ref{eq1}:

  \begin{equation}
  V_H =   \mu_0 [R_\mathrm{0} H + R_\mathrm{S} M_S L(a B)](l/(w t)) I \label{Langevin}
  \end{equation}
  Using the fit of the Hall Voltage dependence with the applied field $H$ (Fig\ref{Fig2}a), it is then possible to get an insight on the mean magnetic moment $\mu $ given by the fit constant  $a=\mu/(\mu_0 k_B T)$. 

 \subsection{Extraordinary Hall Effect}

  For sensor characterization purposes, we also fitted the linear regions around zero field (slope $b=\delta V/\delta B$, EHE contribution) and at saturation (slope $b_0$, OH contribution)(see Eq. \ref{eq1}) as can be seen in Fig\ref{Fig2}b.
  We have used the following set of parameters to characterize each sensor \cite{Giovanni-HallDevices}: the maximum DC current $I$ before any thermal drift is noticeable on the voltage bias $V_b$, the field sensitivity $S_I = b/I$ , the Hall resistance $R_H = V_\mathrm{Hsat}/I$, with $V_\mathrm{Hsat}$ being the intercept of the OH curve, the field saturation $B_\mathrm{sat}= V_\mathrm{Hsat}/b$, and the magnetic field resolution as $B_\mathrm{min}=\delta V_{th}/b$, with the voltage noise limited theoretically by the Johnson noise in the thermal noise frequency range $\delta V_\mathrm{th}=\sqrt{4\pi k_B R T \Delta f}$.

 The Hall constant $R_S$ is the only parameter which cannot be determined directly. It can be obtained either as the intercept of the OH curve: 
$\mu_0 R_S M_S I l/(w t)$  or using the connection between the slope of the EHE curve and the magnetic susceptibility $\chi$: $S_I = R_S \chi l/(w t)$.

 \begin{figure}
 \includegraphics[width=3.8in]{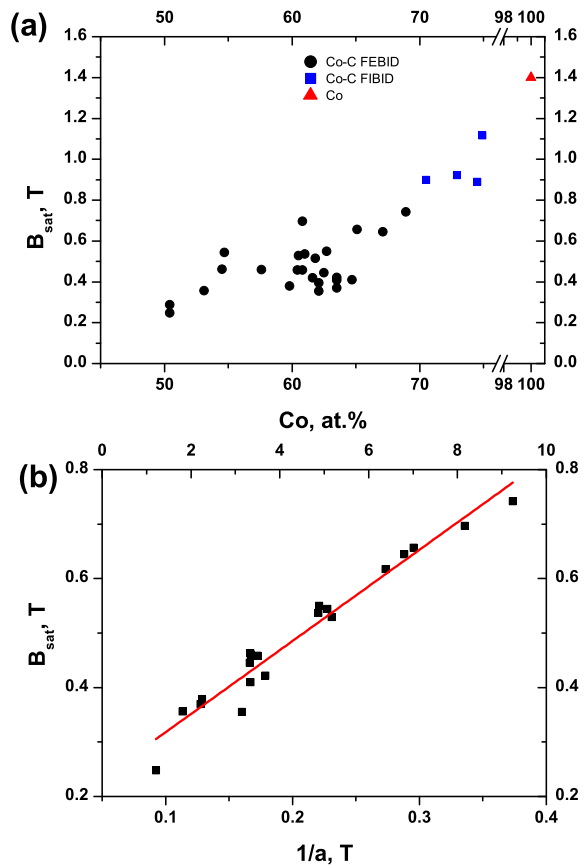}%
 \caption{
 Saturation field $B_\mathrm{sat}$ as a function of: (a) Co atomic concentration, (b) the inverse of the parameter a, from Equation \ref{Langevin}, proportional to the radius $r$ as in Equation \ref{Bsat} \label{Fig3}}
 \end{figure}

\subsection{Saturation Field}
 Figure \ref{Fig2} shows similarity with the magnetization reversal curve in thin magnetic films when the field is applied along the hard axis. However, we are working with superparamagnetic nanoparticles and $B_\mathrm{sat}$ depends on the nanoparticle sizes, rather than film thickness, as encrypted in the Langevin factor (see Fig\ref{Fig3}b):
 \begin{eqnarray}
 \mu H_\mathrm{sat} / k_B T \gg 1,  H_\mathrm{sat} \sim 1/a \sim 1/\mu \sim 1/r^3 \label{Bsat}
 \end{eqnarray}
 As can be seen in Fig\ref{Fig3}a, the FIBID deposits had the higher Co ratios $0.7<x<0.8$ than FEBID deposits, and also higher field saturations. However, since there is no hysteresis, we conclude that this is due to the smaller particle sizes obtained by FIBID.



%

\section{Sensor characterization}
 \begin{figure}
 \includegraphics[width=3.8in]{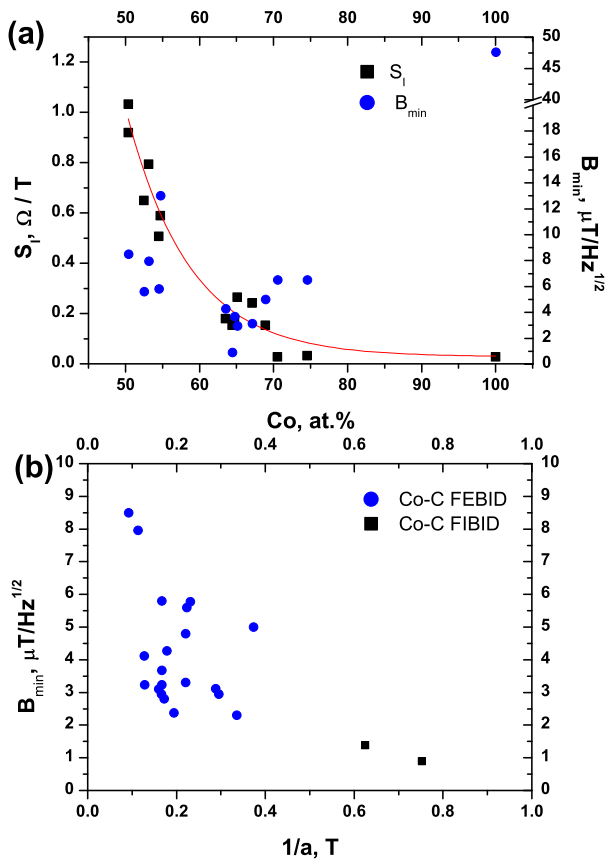}%
 \caption{
 Magnetic field resolution and field sensitivity as a function of: (a) the Co atomic concentration, (b) the inverse of the parameter a \label{Fig4}}
 \end{figure}
The most important parameter for a Hall sensor is the magnetic field resolution, $B_{min}$, defined as the minimum change in the magnetic field that will give a noticeable change in the voltage:
\begin{equation}
 \delta B_{min} = \delta V_\mathrm{th}/ (S_I I_{max}) = \sqrt{4 \pi k_B V_{bias} T \Delta f}/(S_I I_{max}^{3/2})
 \label{Bmin}
 \end{equation}
It depends intrinsically on the maximum current, the field sensitivity and noise spectrum on the Hall voltage contacts. The field sensitivity showed large variation (i.e. $0.04<S_I<1.1$), scaling linearly with the longitudinal resistance and decreasing exponentially with the Co concentrations (Fig\ref{Fig4}a). However, the best magnetic field resolution did not show a clear trend, although the sensors below $5\mu T/Hz^{1/2}$ peaked around 65\% at. Co, where it seems that there was a compromise between low resistivity (high bias current) and a field sensitivity $0.1<S_I<0.4$. Sensors with high $S_I$ also had higher resistance, resulting in an increase in the Johnson noise and a decrease in the bias current. A better understanding may come from the variation of $B_\mathrm{min}$ with the mean magnetic moment, i.e. the particle size - see  Fig\ref{Fig4}b, where we see that the lowest $B_\mathrm{min}$ values were obtained for smallest nanoparticles, for which we expect an enhanced extraordinary Hall resistivity $\rho_\mathrm{EHE}$.


\subsection*{Low field detection}

The 1/f corner frequency (i.e., the frequency where the 1/f is approximately equal to the thermal noise) of our FEBID/FIBID devices is of the order of 100 kHz at the maximum bias current.
For applications in which the field can be modulated above the 1/f corner frequency, the magnetic field resolution B$_\mathrm{min}$ is effectively given by Equation (\ref{Bmin}). However, since the resistance of the realized nanodevices can be as low as 100 Ohms, the thermal noise at the Hall voltage contact can be as low as 1.3 nV/sqrt(Hz).  Preamplifiers having equivalent input noise of the order of 1 nV/sqrt(Hz) or smaller (e.g., Burr-Brown INA103) are, consequently, required to avoid a significant deterioration of the intrinsic magnetic field resolution of the sensor. Otherwise, the possibility to use spinning current techniques to these nanodevices has to be investigated \cite{munter}.

Although the field resolution $B_\mathrm{min}$ of the Co-C sensors is worse than that of semiconducting Hall sensors working at room temperature \cite{Giovanni-HallDevices}, 
the important parameter when detecting localized and close magnetic dipoles (i.e., with dimensions smaller than the Hall sensor active area placed at a distance also smaller than the Hall sensor area) is given by the minimum detectable change in magnetic flux  $\Phi_{min} = B_\mathrm{min} \times A$, where A is the active area of the sensor \cite{Dolabdjian}. Considering the smallest size $200\times200$nm$^2$, the best magnetic flux resolution yields $\phi_{min} = 2\times10^{-5} \phi_0$, where $\phi_0$ is the flux quantum ($\phi_0 = h/2e$), better than semiconducting Hall sensors \cite{Novoselov}.
\subsection*{}






\section{Conclusions}

In conclusion, one can use EHE to characterize granular Co-C wires.  The relation between the sensor characteristics, $S_I$ and $B_\mathrm{min}$ and the structure is non-trivial, but performances are increased with respect to pure Co with an optimum around 65 at.\% Co.  As grown FEBID/FIBID Hall sensors had a $B_\mathrm{min}\sim 1 \mu T/Hz^{1/2}$ and flux resolutions as low as $2\times10^{-5} \phi_0$ at room temperature.

Because of their intrinsic granular structure (enhanced scattering) and nanoparticle sizes below the superparamagnetic limit (hysteresis free), the Co-C Hall devices represent a good 
alternative for a high sensitivity, low field, nano-sized magnetic sensor. 


\begin{acknowledgments}
We would like to thank Dr. Olga Kazakova at National Physics Laboratory, UK, 
for assistance with the magnetic measurements setup and Prof. Hans Hug at EMPA, CH, for the hints on the superparamagnetic characterization. 
This work was supported by the E.C. FP6 under Project BioNano-Switch, Grant No 043288.
\end{acknowledgments}

\bibliography{biblio}

\end{document}